\documentclass{article}
\usepackage{graphicx}
\usepackage{amscd}
\usepackage{amsfonts}
\usepackage{amssymb}
\usepackage{amsmath}
\begin{document}
\hoffset = -2 truecm \voffset = -2 truecm
\baselineskip=12pt
\title{ \textbf{Current quark mass and nonzero-ness of chiral condensates
in thermal Nambu-Jona-Lasinio model} \footnote{The project supported
by the
National Natural Science Foundation of China} \\
}
\author{\textbf{Bang-Rong Zhou}\\
\textbf{College of Physical Sciences, Graduate University of the Chinese}\\
\textbf{Academy of Sciences, Beijing 100049, China}}
\date{}\maketitle
\begin{abstract}
The effect that the current quark mass $M_0$ may result in
nonzero-ness of chiral condensates is systematically reexamined and
analyzed in a two-flavor Nambu-Jona-Lasinio (NJL) model which
simulates Quantum Chromodynamics (QCD) at temperature $T$ and finite
quark chemical potential $\mu$ without and with electrical
neutrality (EN) condition and at any $T$ and $\mu$ without EN
condition. By means of a quantitative investigation of the order
parameter $m$ indicating the chiral symmetry breaking in the model's
ground states, it is shown that a nonzero $M_0$ is bound to lead to
nonzero quark-antiquark condensates throughout chiral phase
transitions within the frame of the NJL model, no matter whether the
order parameter $m$ varies discontinuously or continuously. In fact,
a complete disappearance of the quark-antiquark condensates are
proven to demand the non-physical and unrealistic conditions $\mu
\,\geq$ or $\gg\, \sqrt{\Lambda^2+M_0^2}$ if $T=0$ and  finite, or
$T\to \infty$ if $\mu<\sqrt{\Lambda^2+M_0^2}$, where $\Lambda$ is
the 3D momentum cut of the loop integrals, the largest physical mass
scale in the NJL model.  Theoretically these results show that when
$M_0$ is included, besides the explicit chiral symmetry breaking
indicated by $M_0$ , different from the chiral limit case, we never
have a complete restoration of dynamical (spontaneous) chiral
symmetry breaking, including after a first order chiral phase
transition at low $T$ and high $\mu$.  In physical reality, it is
argued that these results play a decisive role in the known phase
diagram of the model. It is the nonzero-ness of the quark-antiquark
condensates that leads to the appearance of a critical end point in
the first order phase transition line and the crossover behavior at
high $T$ and/or high $\mu$ cases, rather than a possible tricritical
point and a second order phase transition line. They also provide
the basic reason for that one must consider the interplay between
the chiral and diquark condensates in the research on color
superconductor at zero $T$ and high $\mu$ case. The whole
discussions make us learn how a source term of the Lagrangian (at
present i.e. the current quark mass term) can greatly affect
dynamical behavior of a physical system.
\end{abstract}
\section{INTRODUCTION\label{Intro}}
\indent By means of the  Nambu-Jona-Lasinio (NJL) model \cite{kn:1}
which simulates Quantum Chromodynamics (QCD) , one has made
extensive research on chiral phase transitions at finite temperature
$T$ and finite quark chemical potential $\mu$ (corresponding to
quark matter density)\cite{kn:2, kn:3}. It is found that in the
chiral limit where quarks have zero current masses, the dynamical
(spontaneous) chiral symmetry breaking in vacuum induced by the
quark-antiquark condensates, or say, chiral condensates
$\langle\bar{q}q\rangle$ \cite{kn:1,kn:4,kn:5,kn:6} will be restored
at high $T$ and/or high $\mu$, and this restoration will always be
accompanied with the condensates $\langle\bar{q}q\rangle=0$, no
matter whether the symmetry restoring phase transition is first or
second order \cite{kn:7,kn:8}.\\
\indent On the other hand , when there is a nonzero current quark
mass $M_0$, in the vacuum we will have both the explicit chiral
symmetry breaking indicated by $M_0$ and the dynamical chiral
symmetry breaking induced by the condensates
$\langle\bar{q}q\rangle\neq 0$, thus the order parameter $m$
indicating chiral symmetry breaking will include the contributions
from both $M_0$ and the condensates $\langle\bar{q}q\rangle$.
Usually one views $m$ as a whole order parameter indicating chiral
symmetry breaking and does not distinguish it into $M_0$ and the
contribution from the condensates $\langle\bar{q}q\rangle$. However,
in the process of chiral phase transitions at high $T$ and/or high
$\mu$, seeing that $M_0$ is a fixed parameter, what can actually be
changed is merely the $\langle\bar{q}q\rangle$ sector. Therefore,
the chiral phase transitions practically only depend on variation of
the condensates $\langle\bar{q}q\rangle$, but such variation must
presuppose existence of the current quark mass $M_0$. In the
Lagrangian of the model, the current quark mass $M_0$ corresponds a
source term whose existence, in general, will inevitably affect on
dynamical behavior of the model, in present case, i.e. the change of
the condensates $\langle\bar{q}q\rangle$. In view of this, it is
certainly an significant topic of the change of the chiral
condensates $\langle\bar{q}q\rangle$ and its effect on chiral phase
transitions when a current quark mass $M_0$ exists. To focus our
attention on this topic and to compare the obtained results with the
ones in the chiral limit case where $M_0=0$ will be able to make us
acquire a deeper theoretical insight of chiral phase transitions of
the NJL model.\\
\indent As stated above, in the chiral limit, the condensates
$\langle\bar{q}q\rangle$ could be equal to zeros at some high $T$
and/or high $\mu$, hence the dynamical chiral symmetry breaking will
be restored completely. Now one can ask if a similar result could
appear when the current quark mass $M_0$ exists, i.e. at some high
$T$ and/or high $\mu$, the condensates $\langle\bar{q}q\rangle$
disappear totally thus the dynamical (spontaneous) sector of the
chiral symmetry breaking will be restored completely. If the answer
is no, then what physical effects this will actually bring about?\\
\indent The above questions motivate the research of the present
chapter. Although there have been many work about chiral phase
transitions in a NJL model with a current quark mass, it seems still
to lack a systematical and deep examination of the problem from the
point of view considering the relation between the current quark
mass and the chiral condensates, or say, the one between the
explicit and spontaneous chiral symmetry breaking. The discussions
in present chapter will just focus to this respect. It should be
indicated that in the discussions we will revisit the process of
chiral phase transitions at different conditions and this will
naturally involve, at least qualitatively and partly, some known
results. However, it is essential that we will analyzes all the the
results from the above specific point of view and anticipate to get
a deeper understanding about the mechanism of chiral phase
transitions of the NJL model.\\
\indent In the following discussions , we will use the term
"complete restoration of the dynamical chiral symmetry breaking" to
express the condensates $\langle\bar{q}q\rangle=0$.  However, it
should be remembered that in the case with the current quark mass
$M_0$, different from the chiral limit case , the complete
restoration of dynamical chiral symmetry breaking does not mean
chiral symmetry restoration, because after all we always have
the explicit chiral symmetry breaking indicated by $M_0$.\\
\indent We will explore the above topic by means of a two-flavor NJL
model which simulates QCD in the mean field approximation. The
Lagrangian of the model can be expressed by
\begin{eqnarray}
{\cal L}&=&\bar{q}(i\gamma^{\mu}\partial_{\mu}-M_0)q
+G_S[(\bar{q}q)^2+(\bar{q}i\gamma_5\vec{\tau}q)^2]
\end{eqnarray}%%(1)%%
with the quark Dirac fields $q$ in the $SU_f(2)$ doublet and the
$SU_c(3)$ triplets, i.e.
$$q=\left(
\begin{array}{c}
  u_i\\
  d_i \\
\end{array}
\right),\;\; i=r,g,b\; (\mathrm{three \;\;colors}),
$$
$\vec{\tau}=(\tau_1,\tau_2,\tau_3)$ are the Pauli matrices, $G_S$ is
the four-fermion coupling constants and $M_0$ represents the common
current mass of the $u$ and $d$ quarks. The chiral
$SU_{fL}(2)\otimes SU_{fR}(2)$ flavor symmetry of the Lagrangian (1)
will be broken not only explicitly by the current quark mass$M_0$,
but also assumedly in the vacuum spontaneously by the scalar
quark-antiquark condensates $\langle\bar{q}q\rangle$ formed through
the four-fermion interactions $G_S(\bar{q}q)^2$. The constituent
quark mass i.e. the order parameter indicating chiral symmetry
breaking will be $m=M_0-2G_S\langle\bar{q}q\rangle$. It is seen from
this definition of $m$ that the chiral $\langle\bar{q}q\rangle=0$
will mean that $m=M_0$. The emphasis of research will be put on the
effect of the current quark mass $M_0$ on changes of the chiral
condensates $\langle\bar{q}q\rangle$ and the relevant physical
results, including the interesting problem that if the dynamical
chiral symmetry breaking could be restored completely at a
high $T$ and/or a high $\mu$ when $M_0$ exists.\\
\indent In Sect.\ref{0TnoEN} and \ref{0TyesEN}, we will describe the
model's chiral phase transitions in $T=0$ and high $\mu$ case
respectively without and with electric neutrality (EN) condition and
compare the obtained results. By means of a quantitative analysis of
the locations of the least value points of the effective potential,
we will focus on the changes of the chiral condensates in the phase
transitions and the comparison of the results with the ones in
chiral limit case. We will also find out the conditions in which the
chiral condensates could be reduced to zeros and indicate the
non-physical feature of the conditions. In Sect.\ref{general}, a
general analysis of chiral phase transitions of the model will be
conducted in any $T$ and $\mu$ cases without EN condition. We will
indicate the decisive role of nonzero-ness of the chiral condensates
in the phase diagram and give a general demonstration of the
nonzero-ness of the chiral condensates within the frame of the NJL
model. Finally, in Sect.\ref{conclu} we come to our conclusion.
\section{ZERO $T$ AND HIGH $\mu$ PHASE TRANSITIONS \\WITHOUT
ELECTRICAL NEUTRALITY\label{0TnoEN}}

In this section, we will revisit the chiral phase transitions of the
model at $T=0$ and finite $\mu$ without the EN condition and track
the variations of the order parameter indicating chiral symmetry
breaking as $\mu$ increases.\\
\indent When $T=0$, in the mean field approximation, the effective
potential of the model can be expressed by \cite{kn:9}
\begin{eqnarray}
 &&V(m,\mu)\nonumber\\
 &&=\frac{(m-M_0)^2}{4G_S}-12\int\frac{d^3p}{(2\pi)^3}\left[
 E-\sqrt{\vec{p}^2+M_0^2}+\theta(\mu-E)(\mu-E)\right], \;\;E= \sqrt{\vec{p}^2+m^2}\nonumber \\
 &&=\frac{(m-M_0)^2}{4G_S}-\frac{3}{2\pi^2}\left\{
\left[\Lambda(\Lambda^2+m^2)^{3/2}\right.-\frac{m^2}{2}\left(
\Lambda\sqrt{\Lambda^2+m^2}+m^2\ln\frac{\Lambda+\sqrt{\Lambda^2+m^2}}{m}\right)\right.\nonumber \\
&&\left.\hspace{3.6cm}-(m\rightarrow M_0)\frac{{}}{{}}\right]
 \nonumber \\
 &&\left.
+\theta(\mu-m)\left[ \frac{\mu}{3}\sqrt{\mu^2-m^2}(\mu^2-4m^2)+
\frac{m^2}{2}\left(
\mu\sqrt{\mu^2-m^2}+m^2\ln\frac{\mu+\sqrt{\mu^2-m^2}}{m}\right)
\right] \right\},\nonumber \\
\end{eqnarray}%%(2)%%
where $m$ is the constituent quark mass i.e. the order parameter
indicating chiral symmetry breaking, $\mu=\mu_u=\mu_d$ is the common
chemical potential of the two flavor quarks and $\Lambda$ is the 3D
momentum cutoff of the loop integrals.  We will focus on the least
value points of the effective potential $V(m,\mu)$ which correspond
to the ground states of the model.  In all the following
calculations, $\Lambda$ and the four-fermion coupling constant $G_S$
will be fixed by the formula of the $\pi$ decay constant \cite{kn:9}
\begin{eqnarray}
f^2_{\pi}&=&-i\,12m^2\int\frac{d^4p}{(2\pi)^4}
\frac{\theta(\Lambda^2-\vec{p}^2)}{(p^2-m^2+i\varepsilon)^2} \nonumber \\
   &=& \frac{3m^2}{2\pi^2}\left(
\ln\frac{\Lambda+\sqrt{\Lambda^2+m^2}}{m}-\frac{\Lambda}{\sqrt{\Lambda^2+m^2}}
   \right),
\end{eqnarray}%%(3)%%
Gell-Mann Oakes Renner relation \cite{kn:10}
\begin{equation}
    m^2_{\pi}=-\frac{M_0\langle\bar{q}q\rangle}{f^2_{\pi}}
\end{equation}%%(4)%%
and the vacuum form of the following gap equation from the extreme
value condition $\partial V(m,\mu)/\partial m=0$ expressed by
\begin{eqnarray}
m&=&M_0+\frac{6G_S}{\pi^2}m\left[
\Lambda\sqrt{\Lambda^2+m^2}-m^2\ln\frac{\Lambda+\sqrt{\Lambda^2+m^2}}{m}
\right. \nonumber \\
  &&\left.-\theta(\mu-m)\left(
\mu\sqrt{\mu^2-m^2}-m^2\ln\frac{\mu+\sqrt{\mu^2-m^2}}{m}
  \right)\right].
\end{eqnarray}%%(5)%%
By inputting the experimental value $m_{\pi}$=139.57 MeV
\cite{kn:11}, the phenomenological value $f_{\pi}$=92.4 MeV
\cite{kn:12} and taking the current quark mass
\begin{equation}
 M_0=5.55 \;\mathrm{MeV},
\end{equation}%%(6)%%
we will obtain from Eqs.(3),(4) and the equation (5) with $\mu=0$
(vacuum)
\begin{equation}
\Lambda=636.944 \;\mathrm{MeV},\;\; G_S=5.2866\times 10^{-6}\;
\mathrm{MeV}^{-2}.
\end{equation}%%(7)%%
In the meantime we also get $m=m^0_1=322.39$ MeV which is the only
minimum point of $V(m,\mu)$ at $\mu=0$, since it is easy to verify
that the second derivation of $V(m,\mu)$ over $m$
\begin{eqnarray}
\frac{\partial^2V}{\partial m^2}
&=&\frac{1}{2G_S}+\frac{3}{\pi^2}\left[
-\frac{\Lambda^3}{\sqrt{\Lambda^2+m^2}}+3\,m^2\left(
\ln\frac{\Lambda+\sqrt{\Lambda^2+m^2}}{m}-\frac{\Lambda}{\sqrt{\Lambda^2+m^2}}
\right)
\right. \nonumber \\
  &&\left.+\theta(\mu-m)\left(
\mu\sqrt{\mu^2-m^2}-3\,m^2\ln\frac{\mu+\sqrt{\mu^2-m^2}}{m}
  \right)\right]
\end{eqnarray}%%(8)%%
satisfies
\begin{equation*}
\left.\frac{\partial^2V}{\partial
m^2}\right|_{\mu=0,\,m=m^0_1}=35779.3 \;\mathrm{MeV^2} >0.
\end{equation*}
This means (both explicit and spontaneous) chiral symmetry breaking
in vacuum. When we increase $\mu$, the extreme value points of
$V(m,\mu)$ will be determined by Eqs.(5) and (8). The results have
been shown in Table 1. Based on them we may discuss the chiral
phase transitions of the model at $T=0$ and a finite $\mu$.\\
\newpage
Table 1 Variations of the extreme value points of $V(m,\mu)$ as
rising of the quark chemical potential $\mu$. Denotations $(V2)_i$
and $V_i$ respectively represent the values of $\partial^2V/\partial
m^2$ and $V$ at the extreme value points $m^{\mu}_i \,(i=1,2,3)$.\\ \\
\begin{tabular}{|c|c|c|c|c|c|c|c|c|}
  \hline
  $\begin{array}{c}
     \mu \\
     (\mathrm{MeV})
   \end{array}$
  & 0 & 300 & 337 &$\begin{array}{c}
     339.22 \\
     (\mu_{c1})
   \end{array}$ & 340& 350 &500 &$\begin{array}{c}
     636.97 \\
     (\mu_{c2})
   \end{array}$ \\\hline
  $\begin{array}{c}
     m^{\mu}_1 \\
     (\mathrm{MeV})
   \end{array}$ &322.39 &322.39 &308.56 &301.47 &297.85 &  &  &  \\
  $\begin{array}{c}
     (V2)_1 \\
     (\mathrm{MeV}^2)
   \end{array}$ & $>0$ &$>0$  &$>0$  &$>0$ &$>0$ &  &  &  \\
  $\begin{array}{c}
     V_1/10^8 \\
     (\mathrm{MeV}^4)
   \end{array}$ &  &  &-6.901& -6.915 & -6.921 &  &  &  \\\hline
  $\begin{array}{c}
     m^{\mu}_2 \\
     (\mathrm{MeV})
   \end{array}$ & & &174.11 &211.71 &223.10 &  &  &  \\
  $\begin{array}{c}
     (V2)_2 \\
     (\mathrm{MeV}^2)
   \end{array}$ &  &  &$<0$ &$<0$ &$<0$ &  &  &  \\
  $\begin{array}{c}
     V_2/10^8 \\
     (\mathrm{MeV}^4)
   \end{array}$ &  &  &-6.763& -6.858 & -6.886 &  &  &  \\\hline
  $\begin{array}{c}
     m^{\mu}_3 \\
     (\mathrm{MeV})
   \end{array}$ & & &124.17 &102.37 &97.62 &65.78 &11.12 &$\begin{array}{c}
     5.55 \\
     (M_0)
   \end{array}$  \\
  $\begin{array}{c}
     (V2)_3 \\
     (\mathrm{MeV}^2)
   \end{array}$ &  &  &$>0$ &$>0$ &$>0$ &$>0$ & $>0$ & $>0$ \\
  $\begin{array}{c}
     V_3/10^8 \\
     (\mathrm{MeV}^4)
   \end{array}$ &  &  &-6.768& -6.915 & -6.969 &-7.736&-31.66&-83.38\\
   \hline
\end{tabular} \\ \\
First we note that, if we increase $\mu$ but keep $\mu<
m^0_1=322.39$ MeV, for instance, $\mu=300\;\mathrm{MeV}$, then
$V(m,\mu)$ will have the only minimum point still at $m=m^0_1$, as
shown in the left-third column. This indicates that the chiral
symmetry breaking in vacuum will be maintained. However, if $\mu$
goes up to above $m^0_1$, then the situation will be changed. In the
left-fourth column of Table 1 we give an example of $\mu=337
\;\mathrm{MeV}> m^0_1$. In this case, the gap equation (5) will have
three solutions corresponding to two minimum points
$m=m^{\mu}_1=308.56\;\mathrm{MeV}$, $m^{\mu}_3=124.17\;\mathrm{MeV}$
and a maximal point  $m=m^{\mu}_2=174.11\;\mathrm{MeV}$.\\
\indent Obviously , since $V(m_1^{\mu})<V(m_3^{\mu})$,
$m^{\mu}_1=308.56\;\mathrm{MeV}$ will still be the least value point
of $V(m,\mu)$, though it has been less than
$m_1^0=322.39\;\mathrm{MeV}$ in vacuum. It is found that as $\mu$
increases further, $V(m^{\mu}_1)-V(m_3^{\mu})$ will go up, and
finally it will occur that $V(m^{\mu}_1)>V(m_3^{\mu})$ so that the
least value point of $V(m,\mu)$ transfers from $m^{\mu}_1$ to
$m^{\mu}_3$ discontinuously, as shown in the $\mu=340\;\mathrm{MeV}$
column of Table 1 and this indicates that a first order phase
transition has happened. The critical chemical potential $\mu_{c1}$
may be determined by the condition
\begin{equation*}
    V(m_1^{\mu_{c1}},\mu_{c1})=V(m_3^{\mu_{c1}},\mu_{c1})
\end{equation*}
where $m_1^{\mu_{c1}}$ and $m_3^{\mu_{c1}}$ are respectively the two
solutions of Eq.(5) with $\partial^2V/\partial m^2>0$. From this we
obtain $\mu_{c1}=339.22\; \mathrm{MeV}$ with
$m_1^{\mu_{c1}}=301.47\; \mathrm{MeV}$ and $m_3^{\mu_{c1}}=102.37\;
\mathrm{MeV}$, as shown in the left-fifth column of Table 1. It is
emphasized that after the the first order phase transition, we have
the least value point $m_3^{\mu}\gg M_0$ and this implies that the
quark-antiquark condensates $\langle\bar{q}q\rangle$ remain to have
quite large contribution to $m$, hence the dynamical sector of
chiral symmetry breaking, to quite large extent, has still not been
restored.  As a comparison, we note that in the chiral limit
($M_0=0$), we always have the $V$' corresponding second minimal
point $m^{\mu}_3=0$ \cite{kn:7}, hence the present $m^{\mu}_3\gg
M_0$ can only be attributed to the current quark mass effect. Of
course, due to existence of $M_0$, we could at most have
$m^{\mu}_3=M_0$ after a first order phase transition. Assume that is
the case, then the situation will be a little similar to the chiral
limit case, i.e. through a first order phase transition, the sector
of the dynamical chiral symmetry breaking will be restored
completely, though the explicit chiral symmetry breaking indicated
by $M_0$ remains. Then the least value point $m^{\mu}_3=M_0$ will no
longer change as a further increase of $\mu$. The phase transition
will cease at the first order critical value $\mu=\mu_{c1}$.
However, the result $m^{\mu}_3\gg M_0$ has obviously negated the
above assumption.\\
\indent In fact, as $\mu$ continues to increases, the least value
point $m^{\mu}_3$ of $V(m,\mu)$ will smoothly decrease and at above
some value of $\mu$, it will become the only minimum point of $V(m,
\mu)$ left, as shown in the right-third column. A direct question is
whether it may be expected that as $\mu$ grows up further to some
critical value $\mu_{c2}$, we will have $m_3^{\mu}\rightarrow M_0$,
thus the dynamical chiral symmetry breaking will be restored
completely. To answer this question, we may take $m=M_0$ in Eq.(5)
and obtain the equation to determine $\mu_{c2}$
\begin{equation}
M^2_0\ln\frac{\Lambda+\sqrt{\Lambda^2+M^2_0}}{\mu+\sqrt{\mu^2-M^2_0}}
=\Lambda\sqrt{\Lambda^2+M^2_0}-\mu \sqrt{\mu^2-M^2_0}
\end{equation}%%(9)%%
whose solution is
\begin{equation}
 \mu=\mu_{c2}=\sqrt{\Lambda^2+M^2_0}=636.97 \;\mathrm{MeV},
\end{equation}%%(10)%%
in view of Eqs.(6) and (7), as is shown in the right-first column.
However, it is noted that the resulting "second order" critical
chemical potential $\mu_{c2}$ has exceeded the 3D momentum cut off
$\Lambda$ which should be considered as the reasonable largest mass
scale in this NJL model and is nonphysical. The above result shows
that when the current quark mass $M_0$ exists, within the frame of
the NJL model where $\mu<\Lambda$, the limit $m\to M_0$ or
$\langle\bar{q}q\rangle=0$ is not realizable. This fact implies, on
the one hand, that the dynamical chiral symmetry breaking can not be
restored completely; and on the other hand, that after a first order
phase transition one will get only a smooth crossover behavior of
the order parameter $m$ which, however, can never arrive at the
limit value $M_0$.
\section{ZERO $T$ AND HIGH $\mu$ PHASE TRANSITIONS \\WITH
ELECTRICAL NEUTRALITY\label{0TyesEN}} In the above discussions we
did not consider the electrical neutrality (EN) condition of the
quark matter. In this section we will extend our discussions to the
case with EN condition and examine if similar conclusion can be
reached. It is noted that the quark matter with EN is a more
realistic case where electron must be included in the chemical
equilibrium of weak decays of the quarks. In the electric neutrality
case, the effective potential $V(m,\mu,\mu_e)$ in the mean field
approximation can be expressed by \cite{kn:8}

\begin{eqnarray}
&&V(m,\mu,\mu_e) \nonumber \\
&&=\frac{(m-M_0)^2}{4G_S}-6\int\frac{d^3p}{(2\pi)^3}\left[\!\frac{{}}{{}}\right.
2(E-\sqrt{\vec{p}^2+M_0^2})+\theta(\mu_u-E)(\mu_u-E)\nonumber \\
&&\left.\hspace{4.5cm}+\theta(\mu_d-E)(\mu_d-E)\frac{{}}{{}}
\right]-\frac{\mu_e^4}{12\pi^2},\;\;\;\;E=\sqrt{\vec{p}^2+m^2} \nonumber \\
&&=\frac{(m-M_0)^2}{4G_S}-\frac{3}{4\pi^2}\left\{\left[
2\Lambda(\Lambda^2+m^2)^{3/2}\right.-m^2\left(\Lambda\sqrt{\Lambda^2+m^2}\right.+
m^2\ln\frac{\Lambda+\sqrt{\Lambda^2+m^2}}{m}\left.\!\!\frac{{}}{{}}\right)\right.\nonumber\\
&&\left.\hspace{3.8cm}-(m\rightarrow M_0)\dfrac{{}}{{}}
  \right] \nonumber \\
&&\left.\hspace{3.2cm}+\left(
\theta(\mu_u-m)\left[\!\!\frac{{}}{{}}\right.
\frac{\mu_u\sqrt{\mu_u^2-m^2}}{3}\left(\mu_u^2-4m^2\right)+
\frac{m^2}{2}\left(\mu_u\sqrt{\mu_u^2-m^2}\right.\right.\right.\nonumber\\
&&\left.\left.\hspace{4cm}+m^2\ln\frac{\mu_u+\sqrt{\mu_u^2-m^2}}{m}
\left.\left.\!\!\frac{{}}{{}}\right)\right]+(\mu_u\rightarrow\mu_d)
  \right)\right\}-\frac{\mu_e^4}{12\pi^2}\;,
\end{eqnarray}%%(11)%%
where $\mu=-\partial V/\partial n$ is the quark chemical potential
corresponding to the total quark number density $n$, $\mu_e$ is the
chemical potential of electron and
\begin{equation}
 \mu_u=\mu-\frac{2}{3}\,\mu_e, \;\;
 \mu_d=\mu+\frac{1}{3}\,\mu_e=\mu_u+\mu_e,
\end{equation}%%(12)%%
are respectively the chemical potential of the $u,\,d$ quarks.
The last equality of Eq.(12) is usually referred as beta equilibrium \cite{kn:9} \\
\indent For deriving the electrical neutrality condition, it is
noted that the electrical charge density in the two-flavor quark
matter with electrons is
$$n_Q=\frac{2}{3}\,n_u-\frac{1}{3}\,n_d-n_e
$$
with $n_u$, $n_d$ and $n_e$ denoting respectively the number density
of the $u$, $d$ quark and electron. From it we may obtain
$$
\mu_e=-\frac{\partial V}{\partial n_e}=-\frac{\partial V}{\partial
n_Q}\frac{\partial n_Q}{\partial n_e}=-\mu_Q.
$$
Hence the EN condition will become $n_Q=-\partial V/\partial
\mu_Q=\partial V/\partial\mu_e=0$ and has the following explicit
expression
\begin{eqnarray}
\frac{\partial V}{\partial
\mu_e}&=&\frac{1}{3\pi^2}\left\{2\theta(\mu-\frac{2\mu_e}{3}-m)\left[(\mu-\frac{2\mu_e}{3})^2-m^2\right]^{3/2}\right.\nonumber\\
&&\left.-\theta(\mu+\frac{\mu_e}{3}-m)\left[(\mu+\frac{\mu_e}{3})^2-m^2\right]^{3/2}-\mu_e^3\right\}=0.
\end{eqnarray}%%(13)
Owing to Eq.(13), for a given $\mu$, we will have $\mu_e=\mu_e(m)$.
Hence, in view of the EN condition $\partial V/\partial \mu_e=0$,
the extreme value condition of the effective potential
$V(m,\mu,\mu_e)$ now becomes
\begin{equation*}
    \frac{dV}{dm}=\frac{\partial V}{\partial m}+
    \frac{\partial V}{\partial \mu_e}\frac{\partial \mu_e}{\partial
    m}=\frac{\partial V}{\partial m}=0,
\end{equation*}
then from the first equality of Eq.(11), it is easy to obtain the
gap equation $\partial V/\partial m=0$ with the explicit form
\begin{eqnarray}
  m &=&M_0+\frac{3G_S}{\pi^2}\,m\left\{
2\left(
\Lambda\sqrt{\Lambda^2+m^2}-m^2\ln\frac{\Lambda+\sqrt{\Lambda^2+m^2}}{m}
\right)\right. \nonumber \\
   && \left.-\left[
\theta(\mu_u-m)\left(
\mu_u\sqrt{\mu_u^2-m^2}-m^2\ln\frac{\mu_u+\sqrt{\mu_u^2-m^2}}{m}
\right)+(\mu_u\rightarrow\mu_d)
   \right]\right\}.
\end{eqnarray}%%(14)%%
Owing to $dV/dm=\partial V/\partial m$ in the EN condition, the
second derivative of $V(m,\mu,\mu_e)$ over $m$ may be expressed by
\begin{eqnarray*}
\frac{d^2V}{dm^2}&=& \frac{\partial}{\partial
m}\left(\frac{dV}{dm}\right)+
       \frac{\partial}{\partial\mu_e}\left(\frac{dV}{dm}\right)\frac{\partial\mu_e}{\partial m} \\
   &=&\frac{\partial^2V}{\partial m^2}+
       \frac{\partial^2V}{\partial\mu_e\partial m}\frac{\partial\mu_e}{\partial
       m}.
\end{eqnarray*}
From the EN condition $\partial V/\partial\mu_e=0$ we may have
\begin{equation*}
    \frac{d}{dm}\left(\frac{\partial V}{\partial\mu_e}\right)=
     \frac{\partial^2V}{\partial m\partial \mu_e}+\frac{\partial^2V}{\partial\mu_e^2}
     \frac{\partial\mu_e}{\partial m}=0\,.
\end{equation*}
It leads to
\begin{equation*}
    \frac{\partial\mu_e}{\partial m}=-\frac{\partial^2V}{\partial m\partial
    \mu_e}/\frac{\partial^2V}{\partial\mu_e^2}\,.
\end{equation*}
Hence the second derivative of $V(m,\mu,\mu_e)$ over $m$ becomes
\begin{eqnarray}
\frac{d^2V}{dm^2} &=& \frac{\partial^2V}{\partial m^2}-\left(
\frac{\partial^2V}{\partial m\partial\mu_e}\right)^2/\frac{\partial^2V}{\partial\mu_e^2}
      \nonumber \\
  &=& \frac{1}{2G_S}+\frac{3}{\pi^2}\left\{
3m^2\left(
\ln\frac{\Lambda+\sqrt{\Lambda^2+m^2}}{m}-\frac{\Lambda}{\sqrt{\Lambda^2+m^2}}
\right)-\frac{\Lambda^3}{\sqrt{\Lambda^2+m^2}} \right.\nonumber \\
  &&+\frac{1}{2}\left[
\theta(\mu_u-m)\left(
\mu_u\sqrt{\mu_u^2-m^2}-3m^2\ln\frac{\mu_u+\sqrt{\mu_u^2-m^2}}{m}
\right)+(\mu_u\rightarrow\mu_d) \right] \nonumber \\
  &&\left.+
\frac{m^2\left[2\theta(\mu_u-m)\sqrt{\mu_u^2-m^2}-\theta(\mu_d-m)\sqrt{\mu_d^2-m^2}\right]^2}
{4\theta(\mu_u-m)\mu_u\sqrt{\mu_u^2-m^2}+\theta(\mu_d-m)\mu_d\sqrt{\mu_d^2-m^2}+3\mu_e^2}
  \right\}.
\end{eqnarray}%%(15)%%
Now for a given $\mu$, the extreme value points of $V(m,\mu,\mu_e)$
will be determined by the simultaneous equations (13) and (14), and
except for this, the whole discussions of the chiral phase
transitions under EN condition will be parallel to the ones made in
Sect.\ref{0TnoEN}. In Table 2 we list the variations of the extreme
value points of $V(m,\mu,\mu_e)$ for some selected values of $\mu$
in a successively increasing order. \\ \\
\newpage
Table 2  Variations of the extreme value points of $V(m,\mu,\mu_e)$
as rising of $\mu$ under \\electrical neutrality condition.
Denotations $\mu_{ei}$, $(dV2)_i$ and $V_i\;(i=1,2,3)$ represent
\\respectively the values of $\mu_e$, $d^2V/dm^2$  and $V$ at the
extreme value points
$m^{\mu}_i$.\\ \\
\begin{tabular}{|c|c|c|c|c|c|c|c|c|}
  \hline
  $\begin{array}{c}
     \mu \\
     (\mathrm{MeV})
   \end{array}$ & 0 &300 &345 &$\begin{array}{c}
     345.6 \\
     (\mu_{c1})
   \end{array}$ &346 & 350 & 500 &
  $\begin{array}{c}
     656.87 \\
     (\mu_{c2})
   \end{array}$ \\\hline
  $\begin{array}{c}
     m^{\mu}_1 \\
     (\mathrm{MeV})
   \end{array}$  & 322.39 &322.39 &283.69 &277.81 &272.27 & &  &  \\
  $\begin{array}{c}
     \mu_{e1} \\
     (\mathrm{MeV})
   \end{array}$  & 0 & 0 &23.89 &26.18 &28.23 &  &  &  \\
  $\begin{array}{c}
     (dV2)_1 \\
     (\mathrm{MeV^2})
   \end{array}$ & $>0$ &$>0$  &$>0$  &$>0$ &$>0$ &  &  &  \\
  $\begin{array}{c}
     V_1/10^8 \\
     (\mathrm{MeV^4})
   \end{array}$ &  &  &-6.958& -6.967 & -6.974 &  &  &  \\\hline

  $\begin{array}{c}
     m^{\mu}_2 \\
     (\mathrm{MeV})
   \end{array}$  &  & &189.12 &209.32 &221.93 & &  &  \\
  $\begin{array}{c}
     \mu_{e2} \\
     (\mathrm{MeV})
   \end{array}$  &  &  &52.31 &47.32 &43.97 &  &  &  \\
  $\begin{array}{c}
     (dV2)_2 \\
     (\mathrm{MeV^2})
   \end{array}$ &  &  &$<0$  &$<0$ &$<0$ &  &  &  \\
  $\begin{array}{c}
     V_2/10^8 \\
     (\mathrm{MeV^4})
   \end{array}$ &  &  &-6.928& -6.953 & -6.968 &  &  &  \\\hline

  $\begin{array}{c}
     m^{\mu}_3 \\
     (\mathrm{MeV})
   \end{array}$  &  & &143.58 &130.74 &125.3 &96.37&12.3&$\begin{array}{c}
     5.55 \\
     (M_0)
   \end{array}$  \\
  $\begin{array}{c}
     \mu_{e3} \\
     (\mathrm{MeV})
   \end{array}$  &  &  &62.28 &64.53 &65.54 &70.65 &109.41&143.81 \\
  $\begin{array}{c}
     (dV2)_3 \\
     (\mathrm{MeV^2})
   \end{array}$ &  &  &$>0$  &$>0$ &$>0$ &$>0$  &$>0$  & $>0$ \\
  $\begin{array}{c}
     V_3/10^8 \\
     (\mathrm{MeV^4})
   \end{array}$ &  &  &-6.931& -6.967 & -6.992 &  &  &  \\\hline
\end{tabular} \\ \\

The left-second and left-third column of Table 2 indicate that, and
as has been actually checked, the chiral symmetry breaking in vacuum
will be maintained up to $\mu\leq m^0_1=322.39 \;\mathrm{MeV}$. If
$\mu=345\; MeV>m^0_1$ in vacuum, then the effective potential
$V(m,\mu,\mu_e)$ will show two minimal points $m^{\mu}_1$ and
$m^{\mu}_3$ and a maximal point $m^{\mu}_2$.  It may be seen that in
this case $V_1<V_3$ and this merely implies that the least value
point $m^{\mu}_1$ of $V(m,\mu,\mu_e)$ in vacuum smoothly changes to
a lower value and no phase transition happens. However, if
$\mu=346\; \mathrm{MeV}$, we will have $V_1>V_3$.  This indicates
that the ground state has been transferred discontinuously to the
minimal point ($m^{\mu}_3=125.3 \;\mathrm{MeV}, \; \mu_{e3}=65.54
\;\mathrm{MeV}$) and a first phase transition must have happened.
The critical chemical potential $\mu=\mu^{EN}_{c1}$ should be
determined by the equation
\begin{equation*}
   V(m^{\mu}_1,\mu,\mu_{e1})=V(m^{\mu}_3,\mu,\mu_{e3}),
\end{equation*}
where $(m^{\mu}_1,\mu_{e1})$ and $(m^{\mu}_3,\mu_{e3})$ are
respectively the two minimal points of $V(m,\mu,\mu_e)$. The result
$\mu=\mu^{EN}_{c1}=345.6\, \mathrm{MeV}$ is listed in the left-fifth
column of Table 2.\\
\indent At this point, the smaller minimal value point
$m^{\mu_{c1}^{EN}}_3=130.74\,\mathrm{MeV}\gg M_0$, this again
implies that after the first order phase transition the chiral
condensates $\langle\bar{q}q\rangle$ have not been reduced to zeros,
thus the dynamical chiral symmetry breaking has not restored
completely. It is seen from Table 2 that as $\mu$ continue to go up,
for example, for $\mu>350\;\mathrm{MeV}$, the effective potential
$V(m,\mu,\mu_e)$ will have the only minimal point $m_3^{\mu}$ left
whose values will decrease gradually. Finally mathematically
$m_3^{\mu}=M_0$ can be arrived at
$\mu=\mu_{c2}^{EN}=656.87\,\mathrm{MeV}$ and
$\mu_e=\mu_{e3}=143.81\,\mathrm{MeV}$ which are solutions of
Eqs.(13) and (14) with $m$ replaced by $M_0$, as shown in the
left-last column of Table 2. The smooth change of the quark mass
$m^{\mu}_3$ to its current mass $M_0$ indicates that complete
restoration of the dynamical chiral symmetry breaking, if it is
realizable, seems to be a "second order phase transition". However,
it is again found that the "critical chemical potential"
$\mu_{c2}^{EN}$ is higher than the reasonable largest mass scale
$\Lambda$ of the NJL model thus the condition is nonphysical. In
addition, it also demands a quite large
and unrealistic electrical chemical potential $\mu_e$. \\
\indent A comparison between the results with and without EN
condition is shown in Table 3.\\ \\
\centerline{Table 3 Comparison between the results with and without
EN
condition.}\\ \\
\centerline{\begin{tabular}{|c|c|c|c|}
  \hline
         & $\mu_{c1}$ (MeV) & $m_3^{\mu_{c1}}$ (MeV)& $\mu_{c2}$ (MeV)\\\hline
    EN   & 345.6            & 130.74               & 656.87           \\\hline
  non-EN & 339.22           & 102.37               & 636.97 \\
  \hline
\end{tabular}} \\ \\

It is seen that, in the case with EN condition, the first order
critical quark chemical potential $\mu_{c1}$ and the smaller least
value point $m_3^{\mu_{c1}}$ of $V(m,\mu,\mu_e)$ at the critical
point are bigger than the corresponding ones in the case without EN
condition. The former result is consistent with the conclusion
derived in Ref.\cite{kn:8} that in the EN condition, a first order
phase transition must happen at a larger value of $\mu_u$ or
$\mu=\mu_u+(2/3)\mu_e$ than the one in the non-EN condition. In
addition, with EN, the "second order critical chemical potential"
$\mu_{c2}$ to achieve $m\rightarrow M_0$ is higher than the one
without EN, thus it even more exceeds the largest mass scale
$\Lambda$ of the model.\\
\indent In short, inclusion of the EN condition does not change the
qualitative behavior of chiral phase transitions of the model with
the current quark mass $M_0$. Not only after a first order phase
transition but also in the whole variations of $m$, the chiral
condensates $\langle\bar{q}q\rangle$ are always not equal to zeros
within the frame of the NJL model where $\mu<\Lambda$. Therefore, we
can not have complete restoration of the dynamical chiral symmetry
breaking and finally, instead of talking about a second order phase
transition, only get a crossover behavior of the order parameter $m$
as $\mu$ increases.
\section{A GENERAL ANALYSIS FOR ANY $T$ AND $\mu$ CASE\label{general}}

In this section, we will make a general analysis of the chiral phase
transitions of the model in any $T$ and $\mu$ case. For simplicity
and without loss of generality, we will consider only the case
without EN condition. The effective potential $V(m,T, \mu)$ at
finite $T$ and $\mu$ in the mean field approximation may be
expressed by \cite{kn:9}
\begin{eqnarray}
  V(m,T, \mu)&=&\frac{(m-M_0)^2}{4G_S}-12\int\frac{d^3p}{(2\pi)^3}\left\{
E-\sqrt{\vec{p}^2+M_0^2}\right. \nonumber \\
   &&\left.+T\ln\left[1+e^{-(E-\mu)/T}\right]+T\ln\left[1+e^{-(E+\mu)/T}\right]\right\},
   \;E=\sqrt{\vec{p}^2+m^2}.
\end{eqnarray}%%(16)%%
 The gap equation $\partial V(m,T, \mu)/\partial m=0$ becomes
\begin{eqnarray}
m&=&M_0+\frac{12G_S}{\pi^2}\,m \nonumber \\
   &&\times\int_0^{\Lambda}dp\,\frac{p^2}{E}\left[
1-\frac{1}{e^{(E-\mu)/T}+1}-\frac{1}{e^{(E+\mu)/T}+1}
   \right].
\end{eqnarray}%%(17)%%
The second derivative of $V(m,T, \mu)$ over $m$ may be expressed by
\begin{eqnarray}
\frac{\partial^2V}{\partial
m^2}&=&\frac{1}{2\,G_S}-\frac{6}{\pi^2}\int_0^{\Lambda}dp\;
   \frac{p^4}{E^3}\left[1-\frac{1}{e^{(E-\mu)/T}+1}-\frac{1}{e^{(E+\mu)/T}+1}\right]
   \nonumber\\
   && -\frac{6\,m^2}{\pi^2T}\int_0^{\Lambda}dp\;
   \frac{p^2}{E}\left\{
\frac{e^{(E-\mu)/T}}{\left[e^{(E-\mu)/T}+1\right]^2}+(\mu\to -\mu)
   \right\}.
\end{eqnarray}%%(18)%%
By means of Eqs.(17) and (18) and the parameters given by Eqs.(6)
and (7), we may find out the least value points of $V(m,T,\mu)$. For
a given low T, analogous high $\mu$ chiral transitions to the ones
at zero $T$ as described in Sects. \ref{0TnoEN} and \ref{0TyesEN}
will also appear and the discussions may be conducted similarly.
Therefore, in Table 4 we list only the least value points of
$V(m,T,\mu)$ at the critical chemical potential $\mu$ if a first
order phase transition could happen at a given low $T$, or
otherwise,
the only minimal (least) value point of $V(m,T,\mu)$ at a higher $T$.\\

Table 4 Variations of the least value points of $V(m,T,\mu)$ as
change of $T$ and $\mu$.\\
For a given $T$ and $\mu$, $m_1$ and $m_3$ represent the critical
least value points with $V(m_1)=V(m_3)$ and $m_2$ is the
corresponding maximal value point. $(V2)_i$ and $V_i$ $(i=1,2,3)$
represent respectively the values of $\partial^2V/\partial m^2$ and
$V$ at the points $m_i$.\\

\begin{tabular}{|c|c|c|c|c|c|c|c|c|}
  \hline
  $\begin{array}{c}
     T \\
     (\mathrm{MeV})
   \end{array}
  $ & 0 &20 & 35 &37.1 &37.5 &50 &100 & 100 \\\hline
  $\begin{array}{c}
     \mu \\
     (\mathrm{MeV})
   \end{array}$ & 339.22&335.52 &328.48 &327.27 &327.03&327&327&500 \\\hline
  $\begin{array}{c}
     m_1 \\
     (\mathrm{MeV})
   \end{array}$ & 301.47&277.89 &219.93&191.35&186.23&95.47&44.81 &11.17 \\
  $\begin{array}{c}
     (V2)_1 \\
     (\mathrm{MeV^2})
   \end{array}$ &6920.41&4419.43&493.8 &7.287 & $>0$ & $>0$ & $>0$ & $>0$ \\
  $\begin{array}{c}
     V_1/10^8 \\
     (\mathrm{MeV^4})
   \end{array}$ & -6.915 &-7.100&-7.5222&-7.605 &  &  &  &  \\\hline
  $\begin{array}{c}
     m_2 \\
     (\mathrm{MeV})
   \end{array}$ & 211.71&203.17 &189.32&187.58  &  &  &  & \\
  $\begin{array}{c}
     (V2)_2 \\
     (\mathrm{MeV^2})
   \end{array}$ &-2332.48&-1591.39&-220.27 &-4.605 &  &  &  & \\
  $\begin{array}{c}
     V_2/10^8 \\
     (\mathrm{MeV^4})
   \end{array}$ & -6.858 &-7.074&-7.5217&-7.605 &  &  &  &  \\\hline
  $\begin{array}{c}
     m_3\\
     (\mathrm{MeV})
   \end{array}$ & 102.37& 114.53 &156.79 &182.63 & & & & \\
  $\begin{array}{c}
     (V2)_3 \\
     (\mathrm{MeV^2})
   \end{array}$ &3042.59 &2211.86 &387.72 &9.19 & & & &  \\
  $\begin{array}{c}
     V_3/10^8 \\
     (\mathrm{MeV^4})
   \end{array}$ & -6.915 &-7.100&-7.5222&-7.605 &  &  &  &  \\\hline
\end{tabular} \\ \\

It has been verified that at a given sufficiently low temperature
$T$, as $\mu$ increases, we will first have a first order phase
transition and at the first order critical quark chemical potential
$\mu$, we will always have the least value points of $V(m,T,\mu)$
$m_1>m_3\gg M_0$, as shown in the left-first columns of Table 4.
This implies again that after the first order phase transitions we
have not achieved the condensates $\langle\bar{q}q\rangle=0$ and the
complete restoration of the dynamical chiral symmetry breaking,
similar to the case of $T=0$. \\
\indent On the other hand, it is noted that as rising of $T$, for
instance, from 0 to 37.1 MeV, besides that the first order critical
quark chemical potential $\mu$ will slightly go down (from 339.22
MeV to 327.27 MeV), at the first order critical quark chemical
potential $\mu$, the bigger minimal point $m_1$ will go down (from
301.47 MeV down to 191.35 MeV) and the smaller minimal point $m_3$
will go up (from 102.37 MeV up to 182.63 MeV). This shows that the
interval between $m_1$ and $m_3$ will become smaller and smaller. In
the meantime, the curvature ($\propto |\partial^2V/\partial m^2|$)
at the minimal points $m_1$ and $m_3$ (and also including at the
maximal point $m_2$) will also become smaller and smaller. The final
result will be that the two minimal points $m_1$ and $m_3$ will
merge into a single one so that $V(m,T,\mu)$ will have the mere
minimal point $m_1$ left. This indicates that the first order phase
transitions will end at $T=37.1\; \mathrm{MeV}$ and $\mu=327.27\;
\mathrm{MeV}$. Once $T>37.1 \;\mathrm{MeV}$, the effective potential
$V(m,T,\mu)$ will always have the only minimal (also least value)
point left whose location $m_1$ will smoothly decrease as a further
increase of $T$ and $\mu$, as shown in Table 4. As a result, in the
$T-\mu$ phase diagram of the model, $T=37.1 \;\mathrm{MeV}$ and
$\mu=327.27\; \mathrm{MeV}$ will become a critical end pint i.e. at
this point the first order phase transition line ends. \\
\indent It should be emphasized that for the appearance of the
critical end point, a decisive factor is the nonzero-ness of the
chiral condensates $\langle\bar{q}q\rangle$ i.e. $m_3\gg M_0$ after
a first order phase transition. It is the nonzero-ness of
$\langle\bar{q}q\rangle$ that makes it possible that the two minimal
points $m_1$ and $m_3$ could be moved and finally merge into a
single one. Otherwise, if it is assumed that we may have
$m_3\rightarrow$ the fixed $M_0$ i.e.
$\langle\bar{q}q\rangle\rightarrow 0$ after a first order phase
transition, and furthermore, the only minimal point $m_1$ of
$V(m,T,\mu)$ at a higher $T$ can finally be reduced to $M_0$
(similar to a variation of the order parameter in a second order
phase transition), then we will obtain, similar to the chiral limit
case with $M_0=0$, a tricritical point, rather than a
critical end point. \\
\indent Of course, if we have $m_3\gg M_0$ after a first order phase
transition but assume that as a further increase of $T$ and/or
$\mu$, both $m_3$ and the least value point $m_1$ of $V(m,T,\mu)$ at
a higher $T$ could be reduced to $M_0$, then besides that the
explicit chiral symmetry breaking is always kept in the phase
transitions and the critical end point remains to exist, we will
also get a "second order" phase transition line. However, we will
prove that this situation can not practically happen within the
frame of the NJL model.\\
\indent As is shown in Table 4, for the least value points $m$ of
$V(m,T,\mu)$, we always have $m=m_3>M_0$ after a first order phase
transition or $m=m_1>M_0$ at a higher $T$. In addition, similar to
that indicated in the right-last columns of Table 4, $m$ will
decrease as a further increase of $T$ and $\mu$. Thus a natural
question is that if at some high value of $T$ and $\mu$, the limit
$m\to M_0$, corresponding to total disappearance of the chiral
condensates or equivalently, complete restoration of the dynamical
chiral symmetry breaking, could be achieved within the frame of the
NJL model. For giving the answer of this question, we must solve the
gap equation (17) with $m$ replaced by $M_0$. In this case Eq.(17)
becomes
\begin{eqnarray}
   && \int_0^{\Lambda}dp\,\frac{p^2}{E_0}\left[
1-\frac{1}{e^{(E_0-\mu)/T}+1}-\frac{1}{e^{(E_0+\mu)/T}+1}
   \right]=0,\nonumber \\
  &&\hspace{3cm}E_0=\sqrt{p^2+M_0^2}
\end{eqnarray}%%(19)%%
which contains only the parameters $\Lambda$ and $M_0$, independent
of $G_S$. The problem is reduced to find out the
solutions of Eq.(19) about $T$ and $\mu$.\\
\indent First for a given $T$, consider possible values of $\mu$.\\
\indent We begin with $T=0$. When $T\rightarrow0$, we have
$1/[e^{(E_0+\mu)/T}+1]\rightarrow0$, thus Eq.(19) will be reduced to
Eq.(9) which has the solution $\mu=\sqrt{\Lambda^2+M^2_0}$, as given
in Sect.\ref{0TnoEN}.\\
\indent For a finite $T$, we have numerically solved Eq.(19) and
obtained the results listed in Table 5.\\ \\ \\
\centerline{Table 5 Solutions of Eq.(19) for some finite $T$.}\\ \\
\centerline{\begin{tabular}{|c|c|c|c|c|c|}
  \hline
 $T$ (MeV)   & 0      & 20      & 50      & 100     & 200 \\\hline
 $\mu$ (MeV) & 636.97 & 1102.02 & 1858.53 & 3138.57 & 5781.7 \\
  \hline
\end{tabular}} \\ \\

These data indicate that at these (and believably any) finite $T$,
the resulting values of $\mu$ are always bigger or much bigger than
$\Lambda$, thus are not physically reasonable. \\
\indent On the other hand, we may consider the solutions of Eq.(19)
when $\mu<\sqrt{\Lambda^2+M^2_0}$. In fact, numerical calculations
give $T=4.06\times 10^8$ MeV for all $\mu<\sqrt{\Lambda^2+M^2_0}$.
In view of the working precision of the calculations, this merely
indicates that $T\rightarrow\infty$ is the only solution of Eq.(19)
if $\mu<\sqrt{\Lambda^2+M^2_0}$. However, the result $T\to \infty$
is obviously physically unrealistic.\\
\indent The above results can be given a simple physical
explanation. It is noted that the gap equation (17) at finite $T$
and $\mu$ is identical to
\begin{equation}
    m=M_0-2G_S\langle\bar{q}q\rangle_T,
\end{equation}%%(20)%%
where $\langle\bar{q}q\rangle_T$ is the thermal quark-antiquark
condensates which can be obtained from the quark-antiquark
condensates $\langle\bar{q}q\rangle$ in vacuum through replacement
of the quark propagator by its form in thermal field theory
\cite{kn:13}.\\
\indent In fact, the quark-antiquark condensates
$\langle\bar{q}q\rangle$ in vacuum can be expressed by
\begin{equation}
 \langle\bar{q}q\rangle=N_fN_c\int\frac{d^4p}{(2\pi)^4}
 \mathrm{tr}\frac{i}{\not\! p-m+i\varepsilon}\,,
\end{equation}%%(21)%%
where $N_f$ and $N_c$ are the quark's flavor and color number
respectively. Taking the real-time formalism of thermal field
theory, then in Eq.(21) making the replacement
\begin{eqnarray}
 \frac{i}{\not\! p-m+i\varepsilon}&\rightarrow&
 \cos^2\theta_p\frac{i}{\not\! p-m+i\varepsilon}+
 \sin^2\theta_p\frac{i}{\not\! p-m-i\varepsilon}\,,\nonumber \\
 \sin^2\theta_p &=& \theta(p^0)n(p^0-\mu)+\theta(-p^0)n(-p^0+\mu)\,,\nonumber \\
   n(p^0-\mu)&=&\frac{1}{e^{(p^0-\mu)/T}+1}\,,
\end{eqnarray}%%(22)%%
we may express the thermal quark-antiquark condensates by
\begin{eqnarray}
 \langle\bar{q}q\rangle_T &=& -4N_fN_c m\int\frac{d^4p}{(2\pi)^4}\left[
\frac{i}{p^2-m^2+i\varepsilon}-2\pi\sin^2\theta_p\delta(p^2-m^2)
 \right]\nonumber \\
  &=&  -4N_fN_c\int\frac{d^3p}{(2\pi)^3}\frac{m}{2E}\left[
1-n(E-\mu)-n(E+\mu)
  \right]\nonumber \\
  &=&-\frac{6m}{\pi^2}\int_0^{\Lambda}dp\;\frac{p^2}{E}\left[
1-\frac{1}{e^{(E-\mu)/T}+1}-\frac{1}{e^{(E+\mu)/T}+1}
  \right],
\end{eqnarray}%%(23)%%
where $N_f=2$ and $N_c=3$ have been taken. In view of Eq.(23), we
immediately see that Eq.(20) is just the gap equation (17).\\
\indent Now it is easy to find the difference between $M_0=0$ and
$M_0\neq0$ for Eq.(20). When $M_0=0$, i.e. in the chiral limit,
Eq.(20) will become
\begin{equation*}
    1=-\frac{2\,G_S}{m}\langle\bar{q}q\rangle_T.
\end{equation*}
 It may have the solution $m=0$ at some finite and reasonable $T$ and
$\mu$ which is identical to that the thermal condensates
$\langle\bar{q}q\rangle_T=0$ and indicates a complete restoration of
the dynamical chiral symmetry  breaking. On the other hand, when
$M_0\neq0$, if one also expects such complete restoration of the
dynamical chiral symmetry breaking, then it must have the result
that
\begin{equation*}
    \left.\langle\bar{q}q\rangle_T\right|_{m=M_0}=0.
\end{equation*}
However the proceeding discussions in this section just show that
when the current quark mass $M_0\neq 0$ the above result could not
be attained at a finite $T$ and a reasonable $\mu<\Lambda$ i.e.
within the frame of a NJL model where $\mu,T < \Lambda$. We indicate
that it is just a nonzero $\langle\bar{q}q\rangle_T$ that can have a
continuative smooth decrease as rising of $T$ and $\mu$ thus could
lead to the crossover behavior of the order parameter $m$ indicating
chiral symmetry breaking at high $T$ and high $\mu$. Otherwise, if
$m\to M_0$ i.e. the condensates $\langle\bar{q}q\rangle=0$ could be
achieved finally within the frame of the NJL model, then, as stated
above, it will lead to a "second order phase transition line" rather
than the crossover behavior of $m$ in the $T-\mu$ phase diagram.\\
\indent In brief, it is the nonzero-ness of the chiral condensates
induced by the current quark mass $M_0$ which leads to the
appearance of the known critical end point and the crossover
behavior of the order parameter $m$ in the phase diagram of the NJL
model. In view of the simulating feature of the NJL model for QCD,
such phase diagram is very well qualitatively consistent with the
ones obtained based on real QCD dynamics with the current quark mass
\cite{kn:14}.\\
\indent In addition, the condensates $\langle\bar{q}q\rangle\neq0$
at a finite $T,\mu<\Lambda$ demonstrated here will also force one to
consider the interplay between the quark-antiquark condensates
$\langle\bar{q}q\rangle$ and the diquark condensates $\langle
qq\rangle$  when one researches the color superconductors in zero
$T$ and middle $\mu$ case, where $\mu$ has exceeded the first order
critical quark chemical potential $\mu_{c1}$ and the diquark
condensates could be formed. This point has been noticed and
discussed earlier \cite{kn:9,kn:15}, and the analysis made in the
present chapter certainly strengthen theoretical grounds of the
above consideration.
\section{CONCLUSION\label{conclu}}
In this chapter, by means of a quantitative investigation of the
order parameters in the ground states we have reexamined and
analyzed systematically the effect of the current quark mass $M_0$
on nonzero-ness of the chiral condensates $\langle\bar{q}q\rangle$
in chiral phase transitions of a two-flavor NJL model simulating QCD
respectively in the cases of zero $T$ and high $\mu$ without and
with the electrical neutrality condition as well as in the case of
any $T$ and $\mu$ without the electrical neutrality condition. The
nonzero-ness of the chiral condensates $\langle\bar{q}q\rangle$
induced by the current quark mass $M_0$ is found to have a double
meanings: one is that the condensates $\langle\bar{q}q\rangle$ keep
to have quite large values after a first order phase transition at a
low $T$ and a large $\mu=\mu_{c1}$; the other one is that although
at a high $T$ or a large $\mu>\mu_{c1}$, the chiral condensates
$\langle\bar{q}q\rangle$ will smoothly decrease as a further rising
of $T$ and/or $\mu$, they can never be reduced to zeros within the
frame of the NJL model, i.e. in the condition when $\mu, T
<\Lambda$, where the momentum cutoff $\Lambda$ of the loop integrals
must be viewed the largest physical mass scale of the NJL model.\\
\indent For the latter one, in fact, the mathematical solving of the
gap equation indicates that in the case of $T=0$, the order
parameter $m\to M_0$ i.e. the chiral condensates
$\langle\bar{q}q\rangle=0$ could be achieved only at
$\mu=\sqrt{\Lambda^2+M_0^2}$ if the EN condition is not imposed, and
even at a higher $\mu$ if the EN condition is imposed. In the
general case, it has been proven that when the current quark mass
$M_0\neq0$, the limit $m\rightarrow M_0$ can be attained only in the
conditions that $\mu \,\geq$ or $\gg\, \sqrt{\Lambda^2+M_0^2}$ if
$T=0$ and finite, or $T\to \infty$ if $\mu <\sqrt{\Lambda^2+M_0^2}$.
Obviously, these conditions are all
non-physical and unrealistic.\\
\indent The above results indicate that once the current quark mass
$M_0$ exists, it is impossible to achieve complete restoration of
the dynamical chiral symmetry breaking within the frame of the NJL
model. This includes both after a first order phase transition and
in the whole process of chiral phase transitions. Theoretically this
reflects a close and deep interconnection between the explicit and
the dynamical (spontaneous) chiral symmetry breaking.\\
\indent The nonzero-ness of the chiral condensates
$\langle\bar{q}q\rangle$ plays a decisive role in the structure of
the phase diagram of the NJL model with the current quark mass
$M_0$. It is just the nonzero-ness that leads to the appearance of a
critical end point rather than a tricritical point, and a crossover
behavior of the order parameter at high $T$ and $\mu$ rather than a
second order phase transition line in the known phase diagram of the
NJL model. In addition, it also gives a strong theoretical grounds
for the point of view that one must consider the interplay between
the quark-antiquark condensates $\langle\bar{q}q\rangle$ and the
diquark condensates $\langle qq\rangle$ when researching the color
superconductor at zero $T$ and middle $\mu$.\\
\indent To sum up, the analysis made in present chapter of the
nonzero-ness of the chiral condensates induced by current quark mass
and its physical effects may certainly deepen our theoretical
understanding of the mechanism of chiral phase transitions of the
NJL model. It is also an excellent example from which one can
understand how a source term of the Lagrangian can greatly affect
dynamical behavior of a physical system.

\end{document}